\documentclass[conference]{IEEEtran}
\IEEEoverridecommandlockouts
\usepackage{amsmath,amssymb,amsfonts}
\usepackage{algorithmic}
\usepackage{graphicx}
\usepackage{textcomp}
\usepackage{xcolor}
\usepackage[T1]{fontenc}
\usepackage[utf8]{inputenc}
\usepackage{cite}
\usepackage{amsmath,amssymb,amsfonts}
\usepackage{booktabs}
\usepackage{multirow}
\usepackage{array}
\usepackage{url}
\usepackage{balance}
\usepackage{makecell}
\usepackage{threeparttable}

\def\BibTeX{{\rm B\kern-.05em{\sc i\kern-.025em b}\kern-.08em
    T\kern-.1667em\lower.7ex\hbox{E}\kern-.125emX}}
\begin{document}

\title{Bayesian and Classical Feature Ranking for Interpretable BLDC Fault Diagnosis
\thanks{Partially funded by the European Union. Views and opinion expressed are however those of the author(s) only and do not necessarily reflect those of the European Union or Europe’s Rail Joint Undertaking. Neither the European Union nor the granting authority can be held responsible for them. The project FP3-IAM4Rail is supported by the Europe's Rail Joint Undertaking and its members. Partially Funded by AGH's subvention for scientific research.}
}

\author{
\IEEEauthorblockN{Waldemar Bauer}
\IEEEauthorblockA{
\textit{Department of Automatic Control \& Robotics}\\
\textit{AGH University of Krakow}\\
Krakow, Poland \\
0000-0002-8543-0995\\
bauer@agh.edu.pl}
\and
\IEEEauthorblockN{Jerzy Baranowski}
\IEEEauthorblockA{
\textit{Department of Automatic Control \& Robotics}\\
\textit{AGH University of Krakow}\\
Krakow, Poland \\
0000-0003-3313-581X\\
jb@agh.edu.pl}
}

\maketitle

\begin{abstract}
This paper compares Bayesian and classical feature ranking methods for interpretable fault diagnosis of brushless DC (BLDC) motors. Two Bayesian approaches, spike-and-slab and ARD logistic ranking, are evaluated against three classical baselines on a public BLDC benchmark in binary and multiclass settings using current-based, rotational-speed-based, and combined feature sets. The strongest overall results are obtained for the combined representation. In binary classification, ReliefF achieves the highest balanced accuracy of 0.923, while ARD logistic and spike-and-slab remain very close at 0.919 and 0.920 with much smaller subsets ($k=5$). In multiclass classification, ARD logistic performs best for the combined variant with balanced accuracy 0.914, followed closely by LASSO (0.913) and spike-and-slab (0.912). The results show that Bayesian ranking is particularly competitive for current-only and combined descriptors, while ReliefF remains especially effective for speed-based ranking. Because the benchmark consists of short segmented observations from a limited number of experimental conditions, the findings are interpreted primarily as benchmark-specific evidence rather than strong claims of fault generalization.
\end{abstract}


\section{Introduction }
Feature ranking is an important component of interpretable fault diagnosis, especially when models are built from handcrafted time- and frequency-domain descriptors extracted from electrical and mechanical signals. In such settings, compact subsets of informative features may improve predictive performance, reduce redundancy, and support physical interpretation. Classical methods such as ReliefF, mRMR, and LASSO remain standard baselines for this purpose \cite{kononenko1994relief,peng2005mrmr,tibshirani1996lasso}, whereas Bayesian approaches provide an attractive alternative by expressing relevance through posterior uncertainty and hierarchical shrinkage rather than a single deterministic score \cite{george1993gibbs,mackay1995probable,liu2026vard}.

The present study considers feature ranking for BLDC fault diagnosis on the public DUDU-BLDC benchmark \cite{11363866,baranowski_2025_15522163}. This benchmark is compact and consists of short segmented observations derived from a limited number of experimental conditions, so segment-level evaluation may overestimate generalization. For this reason, the paper emphasizes comparative ranking behavior, subset compactness, and stability rather than strong claims about universal fault discrimination performance.

The work is also closely related to our earlier studies on the same benchmark. Previous results established the dataset itself \cite{11363866,baranowski_2025_15522163}, reported Gaussian Naive Bayes baselines \cite{11150867}, and showed that Bayesian fusion of current and speed information can improve multiclass fault recognition \cite{11363817}. In contrast, the present paper does not introduce a new benchmark or a new fusion architecture, but addresses an open question for this dataset: whether Bayesian feature ranking can provide more stable and interpretable subsets than widely used classical ranking methods.

The main contribution of the paper is a comparative study of two Bayesian ranking methods, spike-and-slab and ARD logistic ranking, against three classical baselines, ReliefF, mRMR, and LASSO, on the DUDU-BLDC benchmark in both binary and multiclass settings. The comparison is carried out across current-only, speed-only, and combined feature groups, and evaluates predictive performance, subset compactness, and ranking stability under limited and potentially dependent observations. In this way, the paper contributes an uncertainty-aware analysis of feature relevance for BLDC diagnostics rather than a new dataset or a new classification architecture.

The remainder of the paper is organized as follows. Section II briefly describes the benchmark dataset and diagnostic tasks, Section III introduces the ranking methods, Section IV presents the experimental protocol, Section V reports the results, and Section VI discusses the findings and concludes the paper.

\section{Benchmark Dataset and Diagnostic Tasks}
\subsection{Benchmark Dataset}
The experiments are conducted on a public BLDC diagnostic benchmark introduced in prior work and cited here through its public dataset release. In the merged data file used in this study, the benchmark contains 184 observations and four balanced diagnostic classes:
\begin{itemize}
    \item \texttt{Healthy},
    \item \texttt{Mech\_Damage},
    \item \texttt{Elec\_Damage},
    \item \texttt{Mech\_Elec\_Damage}.
\end{itemize}

Each observation is represented by handcrafted descriptors derived from two signal domains:
\begin{itemize}
    \item current-related features,
    \item rotational-speed-related features.
\end{itemize}

The merged table contains 13 current features and 13 rotational-speed features, yielding 26 modeled descriptors in total. The detailed acquisition protocol and benchmark construction are not repeated here; instead, the original dataset publication is cited as the primary source of documentation.

\begin{table}[t]
\caption{Summary of diagnostic tasks and feature variants.}
\label{tab:dataset_summary}
\centering
\begin{tabular}{lrrr}
\toprule
Task / Variant & Classes & Samples & Features \\
\midrule
Binary / Current & 2 & 184 & 13 \\
Binary / Speed & 2 & 184 & 13 \\
Binary / Combined & 2 & 184 & 26 \\
Multiclass / Current & 4 & 184 & 13 \\
Multiclass / Speed & 4 & 184 & 13 \\
Multiclass / Combined & 4 & 184 & 26 \\
\bottomrule
\end{tabular}

\end{table}

\subsection{Diagnostic Tasks}
Two diagnostic formulations are considered:
\begin{itemize}
    \item \textbf{Binary classification}: \texttt{Healthy} versus all fault conditions merged into one damaged class,
    \item \textbf{Multiclass classification}: \texttt{Healthy}, \texttt{Mech\_Damage}, \texttt{Elec\_Damage}, and \texttt{Mech\_Elec\_Damage}.
\end{itemize}

These two tasks make it possible to study ranking behavior in both simplified fault detection and more detailed fault-type discrimination scenarios.

\section{Feature Ranking Methods}
Five feature ranking methods are considered in this work: ReliefF, mRMR, LASSO, Bayesian spike-and-slab ranking, and Bayesian ARD logistic ranking. For multiclass classification, the Bayesian methods are implemented in a one-vs-rest scheme and the resulting relevance scores are aggregated across classes.

\subsection{ReliefF}
ReliefF is a distance-based feature ranking method that estimates feature relevance by comparing neighboring observations from the same and different classes \cite{kononenko1994relief,urbanowicz2018reliefreview}. Let $w_j$ denote the relevance weight of feature $j$. At each update step, the method compares a sampled observation with its nearest hits and misses, rewarding features that separate observations from different classes and penalizing features that vary within the same class. In simplified form,
\begin{equation}
w_j \leftarrow w_j - \frac{1}{m}\,\mathrm{diff}(x_j,\mathrm{hit}_j)
+ \frac{1}{m}\,\mathrm{diff}(x_j,\mathrm{miss}_j),
\end{equation}
where $m$ is the number of update steps and $\mathrm{diff}(\cdot,\cdot)$ measures discrepancy in the $j$-th feature.

The final ranking is obtained by sorting the features according to $w_j$ in descending order. Relief-based methods remain attractive because they can capture local class structure without assuming a linear predictive model \cite{le2019stir}.

\subsection{mRMR}
The minimum Redundancy Maximum Relevance (mRMR) method ranks features by balancing their relevance to the class label against redundancy with respect to already selected features \cite{peng2005mrmr,hermo2024fedmrmr}. If $S$ is the current subset and $x_j$ is a candidate feature, then the score can be written as
\begin{equation}
\mathrm{mRMR}(x_j)= I(x_j;c)-\frac{1}{|S|}\sum_{x_k\in S} I(x_j;x_k),
\end{equation}
where $I(\cdot;\cdot)$ denotes mutual information and $c$ is the class label.

The first term promotes predictive relevance, whereas the second penalizes redundancy. Features are selected greedily according to this criterion, and the order of selection induces the final ranking.

\subsection{LASSO}
LASSO performs embedded feature selection through $\ell_1$-regularized estimation \cite{tibshirani1996lasso,fira2025sparsefs}. In the present study, it is used in its classification form through logistic regression with an $\ell_1$ penalty. The model parameters are estimated by minimizing
\begin{equation}
\hat{\boldsymbol{\beta}}
=
\arg\min_{\boldsymbol{\beta}}
\left[
-\ell(\boldsymbol{\beta};\mathcal{D})
+\lambda \|\boldsymbol{\beta}\|_1
\right],
\end{equation}
where $\ell(\boldsymbol{\beta};\mathcal{D})$ is the logistic log-likelihood and $\lambda>0$ is the regularization parameter.

The $\ell_1$ penalty shrinks weak coefficients toward zero, thereby inducing sparsity. The feature ranking is obtained from the absolute values $|\hat{\beta}_j|$, with larger magnitudes interpreted as stronger diagnostic relevance.

\subsection{Bayesian Spike-and-Slab Ranking}
The first Bayesian ranking method is based on a spike-and-slab prior for sparse variable selection \cite{george1993gibbs,menacher2024structuredspike}. Let $\gamma_j\in\{0,1\}$ be a latent inclusion indicator for feature $j$. A standard spike-and-slab specification is
\begin{equation}
\beta_j \mid \gamma_j \sim
(1-\gamma_j)\,\mathcal{N}(0,\tau_0^2)
+
\gamma_j\,\mathcal{N}(0,\tau_1^2),
\qquad \tau_0^2 \ll \tau_1^2,
\end{equation}
with
\begin{equation}
\gamma_j \sim \mathrm{Bernoulli}(\pi).
\end{equation}
The spike component strongly shrinks negligible effects toward zero, while the slab component allows practically relevant coefficients to remain active.

Feature ranking is based on the posterior inclusion probability
\begin{equation}
r_j = \Pr(\gamma_j=1 \mid \mathcal{D}),
\end{equation}
where $\mathcal{D}$ denotes the training data. Larger values of $r_j$ indicate stronger posterior evidence that feature $j$ should be included in the predictive model. In the multiclass setting, one-vs-rest models are fitted and the class-specific scores are aggregated into a single ranking.

\subsection{Bayesian ARD Logistic Ranking}
The second Bayesian ranking method is based on Bayesian logistic regression with automatic relevance determination (ARD) \cite{mackay1995probable,neal1996bayesian,liu2026vard}. In ARD, each coefficient is assigned its own feature-specific shrinkage level, which allows the model to regularize weak predictors more strongly than informative ones. A typical hierarchical specification is
\begin{equation}
\beta_j \mid \alpha_j \sim \mathcal{N}(0,\alpha_j^{-1}),
\end{equation}
where $\alpha_j$ is a feature-specific precision parameter. Larger values of $\alpha_j$ imply stronger shrinkage and thus lower relevance of the corresponding feature.

For binary classification, the conditional observation model is
\begin{equation}
y_i \sim \mathrm{Bernoulli}(p_i),
\qquad
p_i = \sigma(\mathbf{x}_i^\top \boldsymbol{\beta}),
\end{equation}
where $\sigma(z)=(1+e^{-z})^{-1}$ denotes the logistic link. In the multiclass setting, ARD is implemented in a one-vs-rest manner and the resulting relevance scores are aggregated across classes.

To preserve the uncertainty-aware character of the ranking, feature $j$ is scored by the posterior threshold-exceedance probability
\begin{equation}
r_j = \Pr\left(|\beta_j|>\epsilon \mid \mathcal{D}\right),
\end{equation}
where $\epsilon$ is a small relevance threshold. Features with larger $r_j$ values are interpreted as more consistently informative under posterior uncertainty.

\subsection{Ranking Aggregation in the Multiclass Setting}
For the Bayesian methods, multiclass feature ranking is performed using one-vs-rest decomposition. Let $r_j^{(k)}$ denote the relevance score of feature $j$ for the classifier separating class $k$ from the remaining classes. The final multiclass relevance score is defined as
\begin{equation}
\bar{r}_j = \frac{1}{K}\sum_{k=1}^{K} r_j^{(k)},
\end{equation}
where $K$ is the number of classes. Features are then ranked according to $\bar{r}_j$.

\section{Experimental Protocol}

\subsection{Overview}
The experiments are designed to compare feature ranking methods rather than to provide a definitive estimate of out-of-sample fault generalization. For this reason, the evaluation protocol emphasizes consistency of ranking behavior, compactness of selected subsets, and predictive performance under repeated resampling.

All preprocessing, feature ranking, and classifier fitting steps are performed using training data only within each split. Test data are used exclusively for final evaluation. This is intended to prevent direct information leakage from the test partition into feature ranking or model fitting.

\subsection{Feature Sets and Tasks}
Three feature-set variants are considered:
\begin{itemize}
    \item \textbf{Current}: 13 current-derived features,
    \item \textbf{Speed}: 13 rotational-speed-derived features,
    \item \textbf{Combined}: all 26 modeled features.
\end{itemize}

Two diagnostic tasks are evaluated:
\begin{itemize}
    \item \textbf{Binary classification}: \texttt{Healthy} versus all fault conditions merged into one damaged class,
    \item \textbf{Multiclass classification}: \texttt{Healthy}, \texttt{Mech\_Damage}, \texttt{Elec\_Damage}, and \texttt{Mech\_Elec\_Damage}.
\end{itemize}

\subsection{Data Standardization}
For each split, input features are standardized using statistics computed on the training partition only. The same transformation is then applied to the corresponding test partition. Standardization is performed separately for each evaluated feature-set variant.

This step is particularly important for coefficient-based ranking methods such as LASSO, spike-and-slab, and Bayesian ARD logistic regression, because their relevance scores depend on the scale of the predictors.

\subsection{Resampling Strategy}
The main evaluation protocol is based on repeated stratified $K$-fold cross-validation. In the present study, we use
\begin{equation}
K = 5
\end{equation}
and repeat the full procedure across multiple random partitions. 

Stratification preserves class proportions in each fold. This choice is motivated by the compact size of the benchmark and the need to compare ranking methods across many train--test configurations. However, because the benchmark is composed of segmented observations originating from a limited number of underlying experimental conditions, the resulting performance estimates should be interpreted with caution. They are used here primarily for comparative analysis of ranking methods rather than as definitive measures of real-world fault generalization.

\subsection{Ranking-and-Classification Pipeline}
For every resampling split, the following steps are carried out:
\begin{enumerate}
    \item select the training and test partitions,
    \item standardize the training data and transform the test data using training statistics only,
    \item fit the ranking method on the training data,
    \item compute a full ranking of features,
    \item select the top-$k$ features,
    \item train the downstream logistic regression classifier using the selected training features,
    \item evaluate the classifier on the corresponding test partition.
\end{enumerate}

This procedure is repeated independently for each ranking method, each task, each feature-set variant, and each tested value of $k$.

\subsection{Downstream Classifier}
To keep the comparison focused and compact, a single downstream classifier is used throughout the study: logistic regression. This choice provides a simple and interpretable predictive model while avoiding confounding effects introduced by varying the classifier family across experiments.

For binary classification, standard logistic regression is used. For multiclass classification, the downstream classifier is implemented in a one-vs-rest form for consistency with the Bayesian ranking procedures.

\subsection{Feature Ranking Methods in the Multiclass Setting}
For ReliefF, mRMR, and LASSO, multiclass ranking is computed directly using the multiclass labels. For the Bayesian methods, multiclass ranking is implemented through one-vs-rest decomposition. Specifically, one binary ranking model is fitted for each class against the remaining classes, and the resulting class-specific relevance scores are aggregated by averaging:
\begin{equation}
\bar{r}_j = \frac{1}{K}\sum_{k=1}^{K} r_j^{(k)},
\end{equation}
where $K$ is the number of classes and $r_j^{(k)}$ denotes the relevance score of feature $j$ for the $k$-th one-vs-rest problem.

\subsection{Subset Sizes}
Each ranking method produces a full ordered list of features. From this ranking, a sequence of top-$k$ subsets is evaluated. In the implementation, we consider the subset sizes
\begin{equation}
k \in \{3,5,7,10,13\}
\end{equation}
for the 13-feature variants and
\begin{equation}
k \in \{3,5,7,10,13,16,20,26\}
\end{equation}
for the combined feature set.

These values allow us to compare compact and moderately sized subsets without overloading the final presentation of results. In the paper, only the best result over tested values of $k$ and the corresponding subset size are reported for each experimental configuration.

\subsection{Bayesian Inference}
The Bayesian spike-and-slab and ARD ranking models are fitted on the training data only. Posterior inference is initially performed using variational inference in PyMC to make repeated fitting across many resampling splits computationally feasible. If posterior behavior or ranking stability indicates substantial approximation issues, selected configurations may additionally be checked using Hamiltonian Monte Carlo as a diagnostic reference.

For Bayesian ARD logistic ranking, feature $j$ is scored using the posterior threshold-exceedance probability
\begin{equation}
r_j = \Pr\left(|\beta_j|>\epsilon \mid \mathcal{D}\right),
\end{equation}
where $\epsilon$ is a small relevance threshold and $\mathcal{D}$ denotes the training data. In the implementation, $\epsilon$ is defined after feature standardization so that the threshold is comparable across predictors. 

For spike-and-slab ranking, the feature score is given by the posterior inclusion probability
\begin{equation}
r_j = \Pr(\gamma_j=1 \mid \mathcal{D}).
\end{equation}

\subsection{Evaluation Metrics}
The primary predictive metric is balanced accuracy, chosen because it remains informative in both the binary and multiclass settings and is appropriate for class-wise comparison even when effective decision difficulty differs across classes.

To complement predictive performance, two additional aspects are evaluated:
\begin{itemize}
    \item \textbf{Subset compactness}, quantified by the number of selected features $k$ required to achieve the best reported result,
    \item \textbf{Ranking stability}, quantified through agreement of selected top-$k$ subsets across repeated resampling runs.
\end{itemize}

Let $S_a^{(k)}$ and $S_b^{(k)}$ denote the top-$k$ feature subsets obtained in two resampling runs. Their overlap is measured using the Jaccard index,
\begin{equation}
J\!\left(S_a^{(k)},S_b^{(k)}\right)
=
\frac{|S_a^{(k)} \cap S_b^{(k)}|}{|S_a^{(k)} \cup S_b^{(k)}|}.
\end{equation}
The reported stability score is the mean Jaccard agreement over repeated runs for the selected subset size.

\subsection{Interpretation of Results}
Because the benchmark is small and built from segmented observations associated with a limited number of underlying experimental conditions, the reported results should be interpreted primarily as benchmark-specific evidence. In this work, the main objective is therefore not to claim universal fault generalization, but to assess which ranking methods produce the most stable, compact, and uncertainty-aware feature subsets under this constrained setting.

\section{Results}

\subsection{Main Comparative Results}

\begin{table}[t]
\caption{Best binary classification results by feature variant.}
\label{tab:main_results_binary}
\centering
\scriptsize
\setlength{\tabcolsep}{3pt}
\renewcommand{\arraystretch}{1.2}
\begin{threeparttable}
\resizebox{\columnwidth}{!}{%
\begin{tabular}{lccccc}
\toprule
Variant & ARD & SS & LAS & Rel & mRMR \\
\midrule
Current  &
\makecell{0.844\\ \scriptsize $k=3$\\ \scriptsize $J=.923$} &
\makecell{\textbf{0.855}\\ \scriptsize $k=7$\\ \scriptsize $J=.943$} &
\makecell{0.850\\ \scriptsize $k=5$\\ \scriptsize $J=.947$} &
\makecell{0.841\\ \scriptsize $k=10$\\ \scriptsize $J=.949$} &
\makecell{0.836\\ \scriptsize $k=13$\\ \scriptsize $J=1.000$} \\

Speed    &
\makecell{0.870\\ \scriptsize $k=7$\\ \scriptsize $J=.839$} &
\makecell{0.865\\ \scriptsize $k=7$\\ \scriptsize $J=.822$} &
\makecell{0.867\\ \scriptsize $k=5$\\ \scriptsize $J=.731$} &
\makecell{\textbf{0.888}\\ \scriptsize $k=3$\\ \scriptsize $J=1.000$} &
\makecell{0.871\\ \scriptsize $k=7$\\ \scriptsize $J=.796$} \\

Combined &
\makecell{0.919\\ \scriptsize $k=5$\\ \scriptsize $J=.855$} &
\makecell{0.920\\ \scriptsize $k=5$\\ \scriptsize $J=.828$} &
\makecell{0.906\\ \scriptsize $k=10$\\ \scriptsize $J=.614$} &
\makecell{\textbf{0.923}\\ \scriptsize $k=20$\\ \scriptsize $J=.942$} &
\makecell{0.908\\ \scriptsize $k=10$\\ \scriptsize $J=.773$} \\
\bottomrule
\end{tabular}%
}
\begin{tablenotes}[flushleft]
\footnotesize
\item Entries report balanced accuracy, subset size $k$, and Jaccard stability $J$. ARD: ARD logistic, SS: Spike-and-Slab, LAS: LASSO, Rel: ReliefF.
\end{tablenotes}
\end{threeparttable}
\end{table}

\begin{table}[t]
\caption{Best multiclass classification results by feature variant.}
\label{tab:main_results_multiclass}
\centering
\scriptsize
\setlength{\tabcolsep}{3pt}
\renewcommand{\arraystretch}{1.2}
\begin{threeparttable}
\resizebox{\columnwidth}{!}{%
\begin{tabular}{lccccc}
\toprule
Variant & ARD & SS & LAS & Rel & mRMR \\
\midrule
Current  &
\makecell{0.859\\ \scriptsize $k=7$\\ \scriptsize $J=.828$} &
\makecell{\textbf{0.861}\\ \scriptsize $k=7$\\ \scriptsize $J=.877$} &
\makecell{0.858\\ \scriptsize $k=7$\\ \scriptsize $J=.927$} &
\makecell{0.839\\ \scriptsize $k=13$\\ \scriptsize $J=1.000$} &
\makecell{0.839\\ \scriptsize $k=13$\\ \scriptsize $J=1.000$} \\

Speed    &
\makecell{0.825\\ \scriptsize $k=7$\\ \scriptsize $J=.896$} &
\makecell{0.805\\ \scriptsize $k=10$\\ \scriptsize $J=.878$} &
\makecell{0.825\\ \scriptsize $k=3$\\ \scriptsize $J=.810$} &
\makecell{\textbf{0.870}\\ \scriptsize $k=5$\\ \scriptsize $J=1.000$} &
\makecell{0.818\\ \scriptsize $k=7$\\ \scriptsize $J=.768$} \\

Combined &
\makecell{\textbf{0.914}\\ \scriptsize $k=20$\\ \scriptsize $J=.863$} &
\makecell{0.912\\ \scriptsize $k=20$\\ \scriptsize $J=.844$} &
\makecell{0.913\\ \scriptsize $k=10$\\ \scriptsize $J=.833$} &
\makecell{0.905\\ \scriptsize $k=26$\\ \scriptsize $J=1.000$} &
\makecell{0.905\\ \scriptsize $k=26$\\ \scriptsize $J=1.000$} \\
\bottomrule
\end{tabular}%
}
\begin{tablenotes}[flushleft]
\footnotesize
\item Entries report balanced accuracy, subset size $k$, and Jaccard stability $J$. ARD: ARD logistic, SS: Spike-and-Slab, LAS: LASSO, Rel: ReliefF.
\end{tablenotes}
\end{threeparttable}
\end{table}

Tables~\ref{tab:main_results_binary} and \ref{tab:main_results_multiclass} summarize the best balanced accuracy, the corresponding subset size, and the ranking stability for all task and feature-set variants. In the binary setting, the combined feature set provides the strongest overall performance. The best binary result for the combined variant is obtained by ReliefF with a balanced accuracy of 0.923, while ARD logistic and spike-and-slab achieve nearly identical performance of 0.919 and 0.920, respectively, using much smaller subsets ($k=5$ for both Bayesian methods versus $k=20$ for ReliefF). For the current-only binary task, spike-and-slab achieves the highest balanced accuracy of 0.855, whereas for the speed-only binary task ReliefF performs best with 0.888 and a compact 3-feature subset.

In the multiclass setting, the combined feature representation again yields the strongest results. ARD logistic achieves the highest balanced accuracy for the combined variant (0.914), followed very closely by LASSO (0.913) and spike-and-slab (0.912). For the current-only multiclass task, spike-and-slab performs best (0.861), narrowly ahead of ARD logistic (0.859) and LASSO (0.858). In contrast, the speed-only multiclass task is dominated by ReliefF, which reaches 0.870 and also exhibits perfect stability. Overall, the Bayesian methods are most competitive in the current-only and combined variants, whereas ReliefF remains particularly strong for speed-based ranking.

The stability values indicate that highly accurate rankings are not always the most stable. For example, in the binary current-only setting mRMR attains perfect stability but the lowest balanced accuracy among the compared methods, while in the multiclass combined setting ReliefF and mRMR are perfectly stable but slightly less accurate than ARD logistic. This confirms that predictive quality and ranking stability should be interpreted jointly rather than separately.

\subsection{Compact Performance Summary}
\begin{figure}[t]
\centering
\includegraphics[width=0.95\columnwidth]{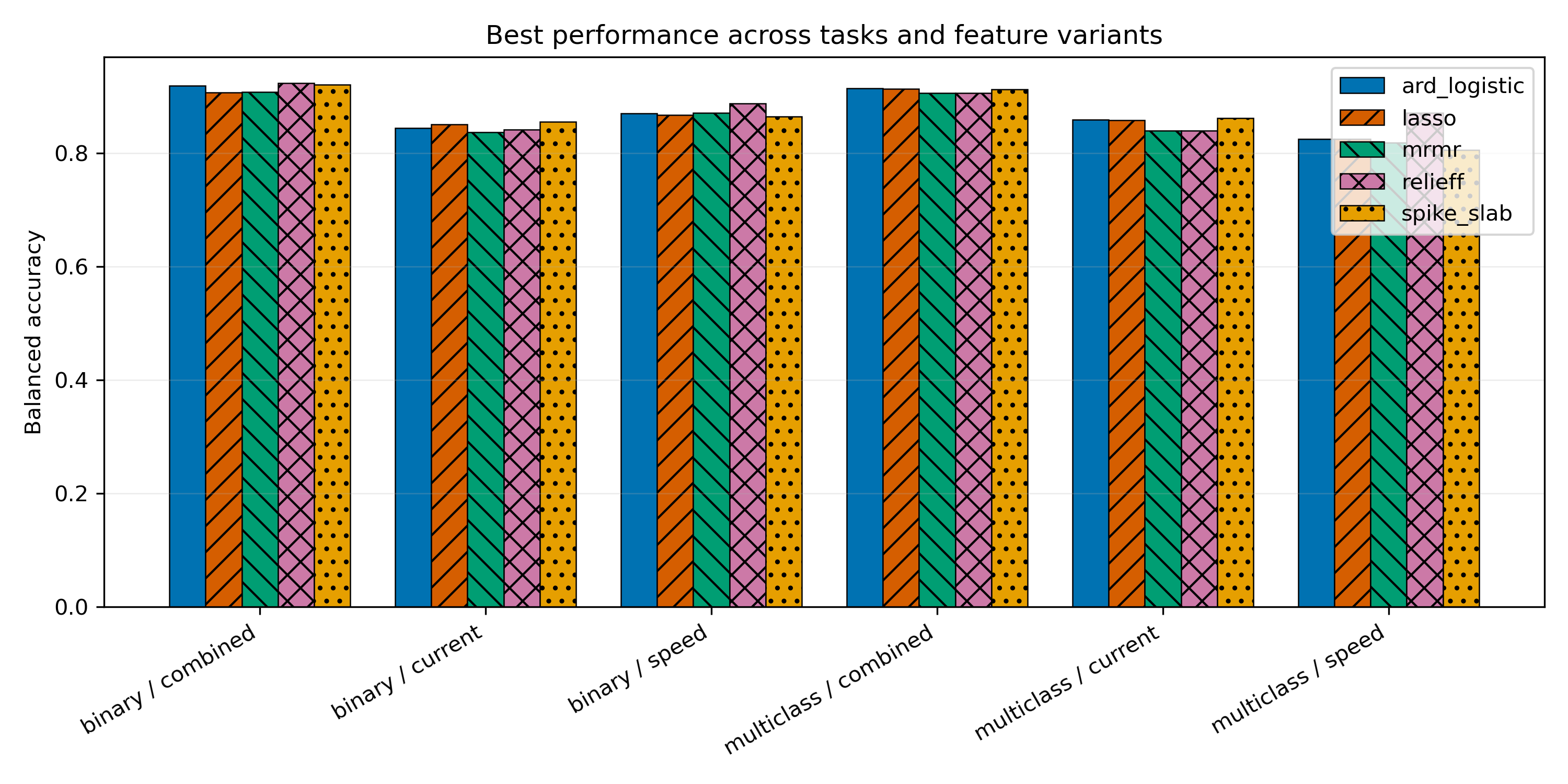}
\caption{Compact comparison of the best predictive performance obtained by all ranking methods across tasks and feature variants.}
\label{fig:performance_summary}
\end{figure}

Figure~\ref{fig:performance_summary} confirms that the combined representation is the strongest overall in both binary and multiclass settings. The advantage of combining current and rotational-speed descriptors is particularly visible in the multiclass problem, where the combined variant consistently outperforms the current-only and speed-only alternatives for most methods. Among the single-domain representations, speed features are generally stronger in the binary task, whereas current-only features become more competitive in multiclass classification when Bayesian ranking is used.

\subsection{Feature-Level Agreement}
\begin{figure}[t]
\centering
\includegraphics[width=0.95\columnwidth]{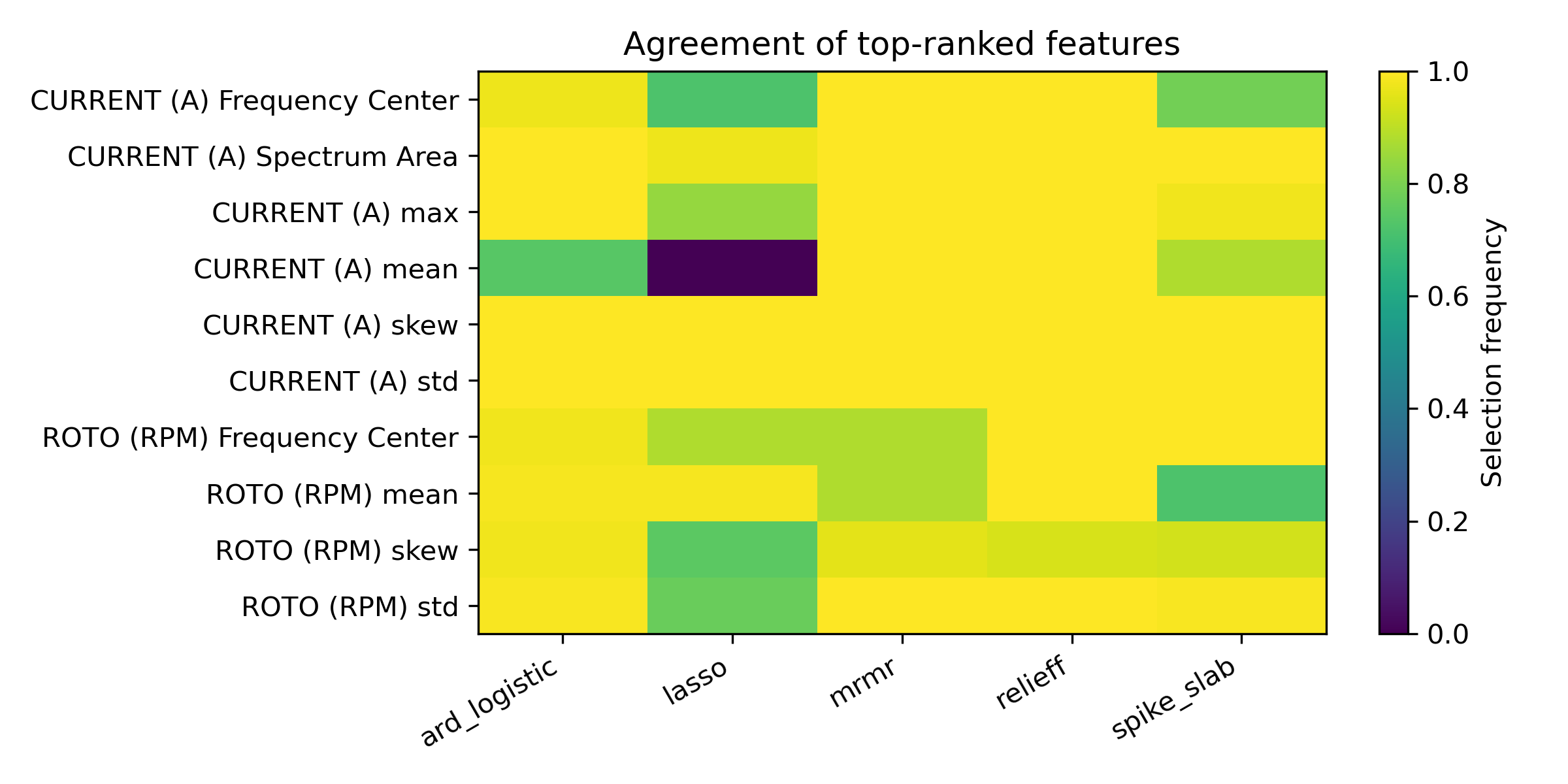}
\caption{Summary of top-ranked features and agreement between Bayesian and classical methods.}
\label{fig:feature_agreement}
\end{figure}

Figure~\ref{fig:feature_agreement} shows that several descriptors are selected consistently across methods. In particular, strong agreement is visible for \texttt{CURRENT (A) Frequency Center}, \texttt{CURRENT (A) Spectrum Area}, \texttt{CURRENT (A) max}, \texttt{CURRENT (A) skew}, as well as \texttt{ROTO (RPM) Spectrum Area}, \texttt{ROTO (RPM) mean}, and \texttt{ROTO (RPM) std}. This pattern suggests that both statistical and spectral descriptors contribute to fault discrimination, with rotational-speed features playing a particularly prominent role in the most accurate subsets.

\section{Discussion and Conclusions}

The reported results show that the combined feature representation is the strongest overall for both diagnostic tasks. In the binary setting, the highest balanced accuracy is obtained by ReliefF for the combined variant (0.923), but the two Bayesian methods remain very competitive, reaching 0.919 for ARD logistic and 0.920 for spike-and-slab while using substantially smaller subsets ($k=5$ in both cases, compared with $k=20$ for ReliefF). In the multiclass setting, the best result for the combined representation is achieved by ARD logistic (0.914), followed very closely by LASSO (0.913) and spike-and-slab (0.912). These results indicate that Bayesian ranking is most competitive for the current-only and combined variants, whereas ReliefF remains particularly strong for the speed-only representation. 

The comparison also shows that predictive quality, subset compactness, and stability should be interpreted jointly. The Bayesian methods are not uniformly the most accurate across all settings, but they often achieve competitive performance with relatively compact subsets and provide an uncertainty-aware view of feature relevance. At the same time, the most stable rankings are not always the most accurate ones. For example, mRMR attains perfect stability in the binary current-only setting but also the lowest balanced accuracy among the compared methods, while in the multiclass combined setting ReliefF and mRMR are perfectly stable but slightly less accurate than ARD logistic. This confirms that stability alone is not sufficient as a selection criterion and should be considered together with predictive performance and subset size. 

The feature-agreement analysis further suggests that a relatively small group of descriptors carries most of the useful diagnostic information. Repeatedly selected features include \texttt{CURRENT (A) Frequency Center}, \texttt{CURRENT (A) Spectrum Area}, \texttt{CURRENT (A) max}, \texttt{CURRENT (A) skew}, as well as \texttt{ROTO (RPM) Spectrum Area}, \texttt{ROTO (RPM) mean}, and \texttt{ROTO (RPM) std}. This pattern is consistent with the broader observation that both statistical and spectral descriptors contribute to discrimination, with rotational-speed features playing a particularly important role in the strongest-performing subsets. 

These findings should nevertheless be interpreted with caution. The benchmark contains 184 observations derived from short segmented recordings and a limited number of underlying experimental conditions, so the reported results should be treated primarily as benchmark-specific evidence rather than definitive fault signatures transferable to all BLDC operating scenarios. Within this constraint, however, the study shows that Bayesian ranking methods constitute meaningful alternatives to classical baselines: ARD logistic provides the strongest overall Bayesian performance, while spike-and-slab remains competitive and offers a complementary sparse relevance perspective. Overall, the paper contributes a comparative and uncertainty-aware view of feature ranking for BLDC diagnostics rather than a new dataset or a new fusion architecture. Future work should focus on stricter evaluation protocols, validation on additional diagnostic datasets, and extension toward more advanced probabilistic relevance models. 

\balance
\bibliographystyle{IEEEtran}
\bibliography{references}

\end{document}